\documentclass[11pt,a4paper,english,nofootinbib,superscriptaddress]{revtex4}
\usepackage{lmodern}

\usepackage[T1]{fontenc}
\usepackage[latin9]{inputenc}
\usepackage{babel}
\usepackage{color}
\usepackage{amsmath}
\usepackage{graphicx}
\usepackage{amssymb}
\usepackage{esint}
\usepackage[unicode=true, pdfusetitle,
 bookmarks=true,bookmarksnumbered=false,bookmarksopen=false,
 breaklinks=false,pdfborder={0 0 1},backref=false,colorlinks=false]
 {hyperref}
\usepackage{amsmath}
\usepackage{amssymb}
\def\barra#1{\not \!#1}
\def\b{\begin{equation}} \def\e{\end{equation}}
\def\bd{\begin{displaystyle}} \def\ed{\end{displaystyle}}
\def\ba{\begin{array}} \def\ea{\end{array}}

\def\bee{\begin{enumerate}}
\def\eee{\end{enumerate}}

\def\ud{\mathrm{d}}

\def\1{\mbox{I\hspace{-.15em}1}}

\def\R{{\rm I\hspace{-.15em}R}}

\def\C{\hspace{3pt}{\rm l\hspace{-.47em}C}}

\def\b{\begin{equation}}
\def\e{\end{equation}}
\def\bee{\begin{enumerate}}
\def\eee{\end{enumerate}}
\def\ii{\mathrm{i}}

\makeatletter
\usepackage{latexsym}\usepackage{bm}

\makeatother
\begin{document}

\title{Higgs Field and the Massless Minimally Coupled Scalar Field \\in de Sitter Universe}

\author{J.-P. Gazeau}
\email{gazeau@apc.in2p3.fr}
\affiliation{APC, Univ Paris Diderot, Sorbonne Paris Cit\'e, France}

\author{M.V. Takook}
\email{takook@razi.ac.ir} 
\affiliation{Department of Physics, Razi University,
Kermanshah, Iran}
\affiliation{Department of Physics,
Science and Research branch, \\ Islamic Azad University, Tehran,
Iran}

\date{\today}

\begin{abstract}

The massless minimally coupled scalar field in the de Sitter space-time is revisited by using the ambient space formalism. We show how this field could play a similar role to that one the Higgs scalar field holds within the electroweak standard model. With the introduction of a ``local transformation'' for that field, a Lagrangian model of interaction  between the latter and a massless spinor field is made comparable to a gauge theory. In the null curvature limit, the Yukawa potential can be constructed from that Lagrangian.  Finally the one-loop correction of the scalar-spinor interaction is presented, which is free of any infrared divergence.

\end{abstract}

\maketitle
\vspace{0.5cm}
{\it Proposed PACS numbers}: 04.62.+v, 03.70+k, 11.10.Cd, 98.80.H
\vspace{1.5cm}

\section{Introduction} 

Since  the LHC has discovered a (possibly the) Higgs boson \cite{higgsbos12,ellis17}, one  gained some certainty in the existence of the corresponding field $\Phi_{H}$  everywhere in the universe. The interaction of this field with vector and spinor fields results in the appearance of mass for some fields through the Higgs mechanism and the spontaneous symmetry breaking. Whereas the interactions between the various fields are usually described by using a gauge formalism,  the interaction between  Higgs and spinor fields is expressed in terms of a Yukawa type potential, which does not pertain to the gauge theory framework. Establishing a firm theory,  based on a gauge principle, for the interaction between Higgs and spinor fields remains  an important challenge.

In this paper, we explain how the massless minimally coupled (\textit{mmc}) scalar field in de Sitter (dS) space  \cite{alfo87,gareta00} could play a  role similar to the special one that the Higgs field holds in the standard model. One essential step is to show that the \textit{mmc} scalar field can be written in terms of the massless conformally invariant (\textit{mcc}) scalar field and an arbitrary constant five-vector $A_\alpha$ in the so-called dS ambient space formalism \cite{dirac35,ta14}. Then we prove that a non-zero constant term in the two-point function, at large relative distance, appears everywhere in the dS universe and that, at short relative distance, we have the Hadamard behaviour, which  guarantees the observation of the field quanta or Higgs particle.

We know that gauge invariance has two handicaps: (1) appearance of infrared divergence and (2) breakdown of background space-time symmetry. Fixing the gauge and using the Gupta-Bleuler triplet formalism \cite{rideau82,araki85}  allow to circumvent these problems. Similar difficulties are  present in the quantization of the \textit{mmc} scalar field in dS space (see \cite{gareta00} and references therein). Here, we propose a new transformation for the \textit{mmc} scalar field in dS space. For the description of the latter  we use the ambient space formalism, and we reformulate the interaction between scalar and spinor fields. We show that  the Yukawa interaction can be reconstructed with an appropriate fixing of the  parameters in the null curvature limit,.

After fixing our notations in section \ref{notations} and recalling the description of the \textit{mmc} scalar field in the ambient space formalism in section \ref{mmcfield1},  we develop our toy model in section \ref{toymod}. Then we show in section \ref{abgt} that for a specific class of abelian gauge parameters our ``local transformation'' becomes equivalent to the latter. Finally, the one-loop correction of the scalar-spinor interaction is presented in section \ref{onelc}. We discuss our results in section \ref{conclu}.  The appendix \ref{vacone} helps to make clearer some aspects of QFT in dS ambient space formalism.

\section{Notations} 
\label{notations}
In the ambient space formalism, the dS space-time is identified with a 4-dimensional hyperboloid embedded in the 5-dimensional Minkowskian space-time $\mathbb{M}_{1,4}$ with the relation:
     \b \label{dSs} M_H=\{x \in \R^5| \; \; x \cdot x=\eta_{\alpha\beta} x^\alpha
 x^\beta =-H^{-2}\},\;\; \alpha,\beta=0,1,2,3,4, \e
where $\eta_{\alpha\beta}=$diag$(1,-1,-1,-1,-1)$ and $H$ is like a Hubble  parameter. The tangential derivative reads as 
\begin{equation}
\label{tangder}
\partial_\beta^\top =\theta_{\alpha \beta}\partial^{\alpha}=
\partial_\beta + H^2 x_\beta x.\partial\,, 
\end{equation}
where $ \theta_{\alpha \beta}=\eta_{\alpha \beta}+
H^2x_{\alpha}x_{\beta}$
is the transverse projector. The intrinsic nature of $\partial_\beta^\top$ implies $ \partial_\beta^\top H^2 =0$  whereas $\partial_\beta  H^2=2H^4 x_\beta$.
The second-order Casimir operator $Q_0$ of the dS group SO$_0(1,4)$ for the scalar field is written as: \begin{equation}
\label{casQ0}
Q_0=-H^{-2}\partial^\top\cdot\partial^\top=-H^{-2}\square_H\,, 
\end{equation}
where $\square_H$ is the Laplace-Beltrami operator on dS space-time \citep{gareta00}. The following identities are easily proved \cite{ta14}:
\begin{align} \nonumber \left[\partial_\alpha^\top\;,\; x_\beta   \right]\phi(x)&= \theta_{\alpha \beta}\phi(x),\;\;\; \left[\partial_\alpha^\top\;,\; \partial_\beta^\top   \right]\phi(x)= H^2x_\beta \partial_\alpha^\top\phi(x) -H^2x _\alpha\partial_\beta^\top\phi(x)\, , \\
 \label{identities} \left[Q_0 \;,\; x_\beta   \right]\phi(x)& = -4x_\beta\phi(x) -2H^{-2}\partial^\top_\beta \phi(x),\;\;\; \left[Q_0\;,\; \partial_\beta^\top   \right]\phi(x)= 2\partial_\beta^\top\phi(x) +2H^2x_\beta Q_0\phi(x) \, . \end{align}
These identities  imply the  two important relations:
\begin{align} \label{identities2} \left[Q_0\;,\; \partial_\beta^\top+2H^2x_\beta  \right]\phi(x)&= -2\left(\partial_\beta^\top +2H^2x_\beta\right)\phi(x) -2H^2 x_\beta\left(2- Q_0\right)\phi(x)\, , \\
 \label{identity3} \left[Q_0 \;, \; \theta_{\alpha\beta}   \right]\phi(x)&= -8x_\alpha x_\beta\phi(x)-2 \theta_{\alpha\beta}\phi(x) -2x_\beta \partial^\top_\alpha \phi(x)-2x_\alpha\partial^\top_\beta \phi(x)\, , \end{align}
Now, the massless conformally coupled (\textit{mcc}) scalar field, denoted by $\Phi_{\mathrm{mcc}}$, satisfies the field equation: 
\begin{equation}
\label{mcceq}
(\square_H+2H^2)\Phi_{\mathrm{mcc}}=0=(Q_0-2)\Phi_{\mathrm{mcc}} \, .
\end{equation}
By applying the identity (\ref{identities2}) to this field, one obtains:
\begin{equation}
\label{mmceq1}
Q_0\left(\partial^\top_\alpha +2H^2x_\alpha \right)\Phi_{\mathrm{mcc}}=0\, . 
\end{equation} 
Multiplying this equation by an arbitrary constant five-vector $A_\alpha$ ($ \partial_\alpha A_\beta=0=\partial_\alpha^\top A_\beta$), we arrive at:
\b \label{mmcs} Q_0\left(A\cdot\partial^\top +2H^2A\cdot x \right)\Phi_{\mathrm{mcc}}=0\, , \e
which is the key equation for our purpose.

\setcounter{equation}{0} 

\section{Massless minimally coupled scalar field} 
\label{mmcfield1}

Due to its appearance in inflationary models \cite{antoilito86} and in quantum linear gravity, the \textit{mmc} scalar field $\Phi_{\mathrm{mmc}}$ has attracted a considerable attention . It satisfies  \cite{gareta00}
\b \label{mmc0}\square_H \Phi_{\mathrm{mmc}}=0=Q_0\Phi_{\mathrm{mmc}}\, .\e In the process of quantization of the \textit{mmc} scalar field, two problems appear: (1) appearance of infrared divergence and (2) breakdown of background space-time symmetry \cite{allen85,alfo87}. Similar difficulties are met in gauge theory. The so-called Krein space quantization, presented and implemented  in \cite{dere98,gareta00},  overcomes these obstacles by  removing the infrared divergence and preserving the de Sitter invariance while  breaking the analyticity.
In addition, the ambient space formalism allowed us to construct a linear gravity, using a polarization tensor and the \textit{mmc} scalar field \cite{taro12,taro15}: 
${\cal K}_{\alpha\beta}=D_{\alpha\beta}(x,\partial)\Phi_{\mathrm{mmc}}$. For recent similar approaches see \cite{pejrahb16}. In Krein space quantization, the scalar field $\Phi_{\mathrm{mmc}}$  was constructed with intrinsic coordinates whilst the polarization tensor $D_{\alpha\beta}(x,\partial)$ was based on ambient space formalism. This  duality might be viewed as a disadvantage on  formal and computational levels. Adopting  the ambient space formalism represents a neutral position. It avoids some drawbacks or artefacts resulting, on the quantum level,  from a specific choice of intrinsic coordinates.  Moreover, two QFT vacuum states are encountered, the Gupta-Bleuler vacuum (for \textit{mmc} scalar field and linear gravity) and the Bunch-Davies vacuum  (for other fields). 
 
From (\ref{mmcs}) we notice that  $\Phi_{\mathrm{mmc}}$ can be  expressed in terms of  $\Phi_{\mathrm{mcc}}$ as
 \begin{equation}
\label{mmcmcc}
 \Phi_{\mathrm{mmc}}(x)\equiv \left[A\cdot\partial^\top + 2H^2 A\cdot x\right]\Phi_{\mathrm{mcc}}(x)\, ,
\end{equation} 
where we remind that $A^\alpha$ is an arbitrary constant five-vector. The quantum field operator $\Phi_{\mathrm{mcc}}$ is constructed from the Bunch-Davies vacuum state in the ambient space formalism \cite{ta14,brgamo94,brmo96}, as is briefly described in the appendix. Since the scalar field $\Phi_{\mathrm{mmc}}$ can be also constructed from the Bunch-Davies vacuum state by application of \eqref{mmcmcc}, it results a quantum field theory with a unique vacuum state, {\it i.e.}, the Bunch-Davies vacuum, and an analytical two-point function can be obtained. 

Indeed, the Wightman two-point function for the \textit{mcc} scalar field reads \cite{brmo96,chta68}:
  \b \label{stpci2}
{\cal W}_{\mathrm{mcc}}(x,x')=\frac{-H^2}{8\pi^2}\frac{1}{1-{\cal Z}(x,x')+\ii\tau \epsilon(x^0,x'^0)}\, ,\quad \epsilon (x^0-x'^0)=\left\{\begin{array}{clcr} 1&x^0>x'^0
 \\
  0&x^0=x'^0\\  -1&x^0<x'^0\\    \end{array} \right.\, ,\e
where $\tau \rightarrow 0$
and 
\begin{equation}
\label{Zxxp}
{\cal Z}(x,x') =-H^2 x\cdot x'=1+\frac{H^2}{2} (x-x')^2\, ,
\end{equation}
 is the geodesic distance between two points $x$ and $x'$ on the de Sitter hyperboloid. 
Using the equation (\ref{mmcmcc}), the two-point function for the \textit{mmc} scalar field is written as:
\b \label{mth} {\cal W}_{\mathrm{mmc}}(x,x';A) =\left[A\cdot\partial^\top + 2H^2 A\cdot x\right]\left[A\cdot\partial'^\top + 2H^2 A\cdot x'\right]{\cal W}_{\mathrm{mcc}}(x,x') \, .\e 
By using the equations
$$ \partial^\top_\alpha x \cdot x'=\partial^\top_\alpha x_\beta x'^\beta=\theta_{\alpha\beta}x'^\beta= x'_\alpha+H^2x_\alpha x\cdot x',\;\;\; \partial^\top_\alpha \frac{1}{1+H^2x\cdot x'}=-H^2\frac{x'_\alpha+H^2x_\alpha x\cdot x'}{(1+H^2x\cdot x')^2}\, , $$ 
and after some simple calculations, one obtains
 \begin{equation}
\label{Wmmc}
\begin{split}
{\cal W}_{\mathrm{mmc}}(x,x';A)&=\frac{-H^2}{8\pi^2}\times\\
&\frac{({\cal Z}-3)\left[(H^2A\cdot x)^2 +(H^2A\cdot x')^2+H^4A\cdot x\, A\cdot x' \,{\cal Z}\right]+6H^4A\cdot x A\cdot x'-(1-{\cal Z})H^2A\cdot A }{(1-{\cal Z}+\ii\tau\epsilon)^3}\, .
\end{split}
\end{equation}
  
This function is not dS invariant with respect to variables $x$, $x^{\prime}$.  Instead, we have to consider the $A$-labelled family of two-point functions for which the following dS invariance holds:
\begin{equation}
\label{Ainvar}
{\cal W}_{\mathrm{mmc}}(Rx,Rx';RA)= {\cal W}_{\mathrm{mmc}}(x,x';A)\quad \mbox{for all} \ R \in \mathrm{SO}_0(1,4)\, .
\end{equation}  
This means that it is built from a one-particle state which does not transform under a unitary irreducible representation of the dS group, a well-known feature of the \textit{mmc} scalar field (see appendix). 
 On the other hand, ${\cal W}_{\mathrm{mmc}}(x,x';A)$ is left invariant by all dS actions belonging to the maximal compact $K_A \sim$ O(4) subgroup  leaving invariant 
the vector $A$. Furthermore, note  the dilation invariance ${\cal W}_{\mathrm{mmc}}(x,x';\lambda A)= \lambda^2 {\cal W}_{\mathrm{mmc}}(x,x';A)$ for all $\lambda \in \R$.

$A^\alpha$ is a constant five-vector. It is a sort of polarization vector \cite{gaha88} which can be chosen as one of the 5 vectors forming an orthonormal (for the metric \eqref{dSs}) basis of $\R^5$, the latter carrying the fundamental five-dimensional representation of the dS group. The explicit form of the two-point function \eqref{Wmmc} depends on the chosen $A^\alpha$. Its construction involves   the tensor product of two representations of the dS group: $(1)$ the scalar complementary representation related to the \textit{mcc} scalar field \cite{ta14}, and $(2)$ the fundamental five-dimensional  representation \cite{gaha88}. 

 As a first simple example of a choice of orthonormal basis in $\R^5$, one considers the set $\{A^{(l)}\, , \, l=0,1,2,3,4\}$ obeying \cite{ta14}:
\b \label{zpolar}  \sum_{l=0}^4 \sum_{l'=0}^4 A^{(l)}_\alpha A^{(l')}_\beta=\eta_{\alpha\beta},\;\;\; A^{(l)}\cdot A^{(l')}=\eta ^{ll'}\, .\e
With this choice and by
summing the 5 corresponding two-point functions one obtains the constant trivial solution:
\b  W_{mmc}(z,z')= \left[\partial^\top\cdot\partial'^\top +2H^2z\cdot\partial'^\top + 2H^2 z' \cdot \partial^\top+ 4 H^4z\cdot z'\right] W_{mcc}(z,z')=\frac{-H^2}{8\pi^2}\, ,\e
with $W_{mcc}$ being the analytic two-point function of the conformally coupled scalar field (\ref{atpfc}). In this case we have restored the trivial SO$(1,4)$ invariance. The following identities were used:
\b  \sum_{l=0}^4 \sum_{l'=0}^4 A^{(l)}\cdot x\, A^{(l')}\cdot x'= x\cdot x'=-H^{-2}{\cal Z},\;\;\;\;\; \sum_{l=0}^4 \sum_{l'=0}^4 A^{(l)}_\alpha A^{(l')}_\beta \eta^{\alpha\beta}=5\, .\e

As a second example, with the elementary choice 
\b \label{so4inv} A_\alpha\equiv (1,0,0,0,0)\, ,\e
we have the  following O$(4)$ invariant two-point function:
\begin{equation}
\label{ }
{\cal W}_{\mathrm{mmc}}(x,x') =\frac{-H^2}{8\pi^2}
\frac{({\cal Z}-3)\left[(H^2x^0)^2 +(H^2 x'^0)^2+ H^4x^0\,  x'^0 \,{\cal Z}\right]+6 H^4 x^0  x'^0-H^2(1-{\cal Z}) }{(1-{\cal Z}+\ii\tau\epsilon)^3}\, .
\end{equation} 

The 2-point function function \eqref{Wmmc} is free of logarithmic divergence, contrary to    the 2-point function ${\cal W}^{\mathrm{AF}}_{mmc}$ discussed by Allen-Folacci in \cite{alfo87}. As a matter of fact, the expression of the latter has the following logarithmic divergence : 
\b {\cal W}^{\mathrm{AF}}_{mmc}(x,x') \propto \left[\frac{1}{1-{\cal Z}}-\ln (1-{\cal Z})+ \cdots\right]\,.\e

The singularity of the two-point  function \eqref{Wmmc} in the limit $x \longrightarrow x'$ (${\cal Z}= 1$) is similar to the Hadamard behavior of the two-point function. On the other hand, at large ${\cal Z} \sim (x-x^{\prime})^2\sim -x\cdot x^{\prime}\longrightarrow \infty$, the dominant term in  ${\cal W}_{\mathrm{mmc}}$ is 
$$-\frac{H^4}{8\pi^2}\left[\frac{A\cdot x\, A\cdot x' }{ x\cdot x^{\prime}}+  \frac{(A\cdot x)^2 +  (A\cdot x^{\prime})^2}{ (x\cdot x^{\prime})^2}\right]\, . $$
One can prove from this expression that the two-point  function can assume any  value at large ${\cal Z}$. For instance, with the choices $A= (A_0,0,0,0,0)$, $x=H^{-1} (\sinh\psi,0,0,0,\cosh\psi)$, and for fixed $x^{\prime}=H^{-1} (\sinh\psi^{\prime},0,0,0,\cosh\psi^{\prime})$, the above expression reduces at large $-x\cdot x^{\prime}= H^{-2}\cosh(\psi - \psi^{\prime})$, i.e., at large $\psi$,  to
$$
-\frac{A_0^2H^4}{16\pi^2} \left[1+ \cosh\psi^{\prime} + \tanh(\psi-\psi^{\prime})\sinh\psi^{\prime}\right] \sim -\frac{A_0^2H^4}{16\pi^2}\left(1+ e^{2\psi^{\prime}}\right) \ \mbox{as} \ \psi \to \infty \, ,
$$
and, of course, $\psi^{\prime}$ can be made arbitrarily large provided it remains smaller than $\psi$ such that $\psi-\psi^{\prime} \to \infty$. 
 Hence, we get a $A$ dependent  value, which  depends in general on the respective directions of $x$ and $x^{\prime}$ along which their  large separation limit is performed. This behavior of the two-point function at large separation distances appears also for massless spins $\frac{3}{2}$ and $2$ since the latter can be written in terms of the polarization spinor-tensor and a \textit{mmc} scalar field ($ \Psi_\alpha={\cal D}_\alpha (x,\partial)\Phi_{\mathrm{mmc}}, $) \cite{ta14,taro12,fatata14}. For recent similar approaches see \cite{pejrahb16}. As well as the  massive fields, the \textit{mmc} scalar field is not conformally invariant.  Now, we should be aware that the conformal invariance of  the massless spin $\frac{3}{2}$ and $2$ fields is  broken after covariant quantization. This may be a cause of the appearance of mass in a physical theory, similarly to the Higgs mechanism.

 What is really the physical meaning of $A^\alpha$ and how it can be determined? The physical meaning is discussed in appendix \ref{vacone}, and it may be fixed by considering  the interaction cases in the null curvature limit. In the next section we present a toy model for obtaining the interaction Lagrangian.

\setcounter{equation}{0} 
\section{A Toy model} 
\label{toymod}
In the previous section we noticed the arbitrariness in the choice of   $A^\alpha$. We know that all scalar equations  in dS space of the form 
\begin{equation}
\label{ }
\left(Q_0+ \sigma(\sigma + 3)\right)\Phi(x)=0\, ,
\end{equation}
have  a continuous family of simple solutions, named ``de Sitter plane waves''
\begin{equation}
\label{genfctrep}
\Phi(x)=(x\cdot \xi)^{\sigma}\, .
\end{equation}
which  are   indexed by vectors $\xi$ lying in the positive null-cone in $\mathbb{M}_{1,4}$. Those vectors play the r\^ole of a momentum parameter. Thus, in the \textit{mmc\textit{}} case, we have $\sigma= 0$ (constant solution) or $\sigma = -3$. In the \textit{mcc} case, $\sigma =-
 2$  or $\sigma = -1$ and \eqref{genfctrep} generates a complete set of solutions. The values in the interval $-3<\sigma< 0$ label the so-called scalar complementary series  of unitary irreducible representations of the dS group, and  the boundary 
$\sigma = -3$ marks the departure of the scalar discrete series \cite{gasiyou10}. From the ladder relation
\begin{equation}
\label{ladder}
\left(A\cdot\partial^\top -\sigma H^2 A\cdot x\right) (x\cdot \xi)^{\sigma}= \sigma A\cdot \xi\, (x\cdot \xi)^{\sigma-1}
\end{equation} 
we easily understand the meaning of the operator $A\cdot\partial^\top + 2H^2 A\cdot x$ in   \eqref{mmcmcc} which corresponds to $\sigma = -2$. 

Trivially the field equation (\ref{mmc0}) is invariant under the  constant transformation
\begin{equation} \label{sgt}
\Phi_{\mathrm{mmc}} \longrightarrow \Phi_{\mathrm{mmc}}'=\Phi_{\mathrm{mmc}}+\Phi_g\, , \;\; \Phi_g=\mbox{constant}\, .
\end{equation}
Alternatively, from the above observations and trivial linear superposition of solutions, this equation is also invariant under the   transformation 
\begin{equation} \label{sgt2}
\Phi_{\mathrm{mmc}} \longrightarrow \Phi_{\mathrm{mmc}}'=\Phi_{\mathrm{mmc}}+(x\cdot B) ^{-3}\,,
\end{equation}
where $B^\alpha$ is an arbitrary constant five-vector on the null-cone in $\mathbb{M}_{1,4}$. To some extent, we meet a similar situation in gauge theory with the choice of the gauge potential.

Let us construct a toy model of interaction from \eqref{sgt2}. For the sake of simplicity we consider the interaction with a spinor field. We consider a massless conformally invariant spinor field in dS universe. Its action functional reads \cite{bagamota01,ta14}
\b \label{actionspior} S(\psi)=\int \ud\mu(x){\cal L}(\psi)=\int \ud\mu(x)H \psi^\dag \gamma^0\left( -\ii\barra{x}\gamma^\alpha \partial^\top_\alpha+2\ii\right) \psi\, ,\e
where $\ud\mu(x)$ is the invariant volume element on the dS hyperboloid and $\barra{x}= \gamma_\alpha x^\alpha$. The  five $4\times4$ matrices $\gamma^{\alpha}$  satisfy the Clifford conditions \cite{ta14,bagamota01}:
$$\gamma^{\alpha}\gamma^{\beta}+\gamma^{\beta}\gamma^{\alpha}
=2\eta^{\alpha\beta}\, , \qquad
\gamma^{\alpha\dagger}=\gamma^{0}\gamma^{\alpha}\gamma^{0}\,.$$ 
The following representation for the $\gamma$ matrices is well adapted to the dS ambient space formalism \cite{ta14,bagamota01}:
\begin{align}
\nonumber     \gamma^0 &=\left( \begin{array}{clcr} I & \;\;0 \\ 0 &-I \\ \end{array} \right)
      ,\;\;\;\gamma^4=\left( \begin{array}{clcr} 0 & I \\ -I &0 \\ \end{array} \right) \, ,  \\
 \label{gammam}
   \gamma^1   & =\left( \begin{array}{clcr} 0 & \ii\sigma^1 \\ \ii\sigma^1 &0 \\
    \end{array} \right)
   ,\;\;\gamma^2=\left( \begin{array}{clcr} 0 & -\ii\sigma^2 \\ -\ii\sigma^2 &0 \\
      \end{array} \right)
   , \;\;\gamma^3=\left( \begin{array}{clcr} 0 & \ii\sigma^3 \\ \ii\sigma^3 &0 \\
      \end{array} \right)\, , 
\end{align}  
where the $\sigma^i$'s $(i=1,2,3)$ are the Pauli matrices and $I$ is the unit $2\times 2$  matrix.

The action (\ref{actionspior}) is invariant under a global $U(1)$ phase transformation. Let us assume that the spinor field $\psi$, under the ``local transformation'' (\ref{sgt2}),  becomes:
\begin{equation} \label{spgt}
\psi \longrightarrow \psi^{\prime}=U(x,B)\psi=e^{-\ii\Lambda(x,B)}\psi\, ,
\end{equation}
where $\Lambda$ is  an arbitrary function. Now the tangential derivative $\partial^\top_\alpha$ does not commute with  $U$,  $\partial_{\alpha}^\top \psi \longrightarrow \partial_{\alpha}^\top \psi^{\prime} \neq U \partial_{\alpha}^\top \psi $. 
Hence we define the new derivative and transformation $U$ as
\begin{equation} \label{gcd}
D_{\alpha}^\Phi \psi=\left(\partial_{\alpha}^\top+\ii B^\top_\alpha \Phi_{\mathrm{mmc}}\right)\psi\,,\e 
\b \label{u1gts} U=\exp \left[\frac{\ii}{2} (x\cdot B)^{-2}\right]\, ,
\end{equation}
in order to get the transformation:
\begin{equation} 
 D_{\alpha}^\Phi \psi \longrightarrow (D_{\alpha}^\Phi \psi)' =D_{\alpha}^{\Phi'} \psi'= UD_{\alpha}^\Phi \psi\, .
\end{equation}
We see that the function $\Lambda=-\frac{1}{2} (x\cdot B)^{-2}$ is not completely arbitrary whereas $(B_\alpha)$ is a  constant vector in the null-cone. Thus it is not a true gauge transformation. By replacing  in \eqref{actionspior} the tangential derivative with the covariant derivative (\ref{gcd}) we obtain:
$$ S(\psi)=\int \ud\mu(x) \psi^\dag \gamma^0H\left[-\ii \barra{x}\gamma^\alpha \left(\partial_{\alpha}^\top+\ii B^\top_\alpha \Phi_{\mathrm{mmc}}\right)+2\ii\right] \psi\, .$$ is invariant under the following transformations
\b \label{localtran} \Phi_{\mathrm{mmc}}'=\Phi_{\mathrm{mmc}}+(x\cdot B) ^{-3}, \;\;\; \psi'=e^{\frac{\ii}{2} (x\cdot B) ^{-2}}\psi\, .\e
In this case the scalar field $\Phi_{\mathrm{mmc}}$ and the constant vector  $B$ may be considered as the gauge potential and  the generators of this transformation respectively.

Therefore the interaction Lagrangian between the scalar field  $\Phi_{\mathrm{mmc}}$ and the spinor field $\psi$, defined as
\begin{equation}
{\cal L}_{int}=H\psi^\dag \gamma^0 \barra{x}\gamma^\alpha  B_{\alpha}^{\top} \Phi_{\mathrm{mmc}}\psi \, ,
\end{equation}
 can be exactly identified, in the null curvature limit, with the Yukawa interaction type with a convenient choice of the five-vector constant $B^{\alpha}$. It is interesting to note that the arbitrariness of $B$ is fixed in the null curvature limit and in a specific coordinate system, which is  similar to the gauge fixing in the general coordinate transformation. Our  model is also  similar to the definition of the Higgs bundle and its connection \cite{wen14,kawi06}. Exploiting the similarity of the $\Phi_{\mathrm{mmc}}$ with the Higgs field, the null curvature limit must be carefully studied. This similarity lets us think about a possible relation between  $B^{\alpha}$ and  a kind of internal structure of the observed Higgs. 
 
 To conclude this section,  we claim that our toy model can be used for defining the interaction between  scalar and spinor fields via the transformation (\ref{localtran}).

\setcounter{equation}{0} 
\section{Abelian gauge theory} 
\label{abgt}
Let us recall briefly the abelian gauge theory expressed in terms of  ambient space coordinates \cite{rota05}. A massless vector field or gauge vector field satisfies   \citep{gagarota08}:
\b\label{mlvfe} Q_1K_{\alpha}+\partial^\top_\alpha \partial^\top\cdot K=0\, ,\e
where $Q_1K_\alpha= \left(Q_0-2\right)K_{\alpha}+2x_\alpha \partial^\top\cdot K$ is the Casimir operator action on  vector fields.  The corresponding Lagrangian reads $ {\cal L}(K)= K\cdot Q_1K +K\cdot\partial^\top \partial^\top\cdot K$.
By using the equations (\ref{identities}), the following identities are easily proved:
\b \label{identity4} Q_1 \partial^\top_\alpha \phi=\partial^\top_\alpha Q_0\phi,\;\;  Q_1 B^\top_\alpha \phi=B^\top_\alpha \left(Q_0-2 \right)\phi-2 \partial^\top_\alpha B\cdot x \phi\, ,\e
where $B_\alpha$ is a constant five-vector in the null-cone and $B^\top= B+ H^2x\cdot B \,x$.  By noticing  that the field equation (\ref{mlvfe}) is invariant under the  gauge transformation
\b \label{posibili1}  K_\alpha \rightarrow K^g_\alpha=K_\alpha+\partial^\top_\alpha \Lambda\, ,\e 
where $\Lambda(x)$ is an arbitrary scalar field,  one shows that the action (\ref{actionspior}) is invariant under the local  transformation $ \psi \rightarrow \psi^G=e^{-i\Lambda(x)}\psi $ 
if the tangential derivative $\partial^\top_\alpha$ is replaced with the gauge-covariant derivative \cite{ta14,rota05}
$D_{\alpha}^K =\partial_{\alpha}^\top+iK_\alpha$.
 
With the choices $\Lambda\equiv-\frac{1}{2} (x\cdot B) ^{-2}$ and $K_\alpha\equiv B^\top_\alpha\Phi_{\mathrm{mmc}}$, our  model becomes exactly the above Abelian gauge theory:
\b \label{equi} B^\top_\alpha\Phi_{\mathrm{mmc}}'= B^\top_\alpha\Phi_{\mathrm{mmc}}+\partial^\top_\alpha\left[-\frac{1}{2}(x\cdot B) ^{-2}\right]=B^\top_\alpha\Phi_{\mathrm{mmc}}+B^\top_\alpha(x\cdot B) ^{-3}\, . \e
Imposing that the vector field $K_\alpha\equiv B_{\alpha}^{\top}\Phi_{\mathrm{mmc}}$ satisfy the same field equations as the massless vector field (\ref{mlvfe}) leads to the  constraint
\b \label{constraint}  \partial^\top_\alpha B\cdot\partial^\top \Phi_{\mathrm{mmc}} + 2H^2 B\cdot x\partial^\top_\alpha \Phi_{\mathrm{mmc}}=0\, ,\e
for which  we used
\b Q_0B_{\alpha}^{\top} \phi- B_{\alpha}^{\top}Q_0 \phi=-8B\cdot x x_\alpha \phi-2B_{\alpha}^{\top} \phi-2 B\cdot x \partial^\top_\alpha  \phi-2x_\alpha B\cdot \partial^\top \phi\, .\e

The relations $B\cdot K= (B\cdot x)^2 \Phi_{\mathrm{mmc}}$ and $\partial^\top\cdot K=4B\cdot x \Phi_{\mathrm{mmc}}+ B\cdot \partial^\top \Phi_{\mathrm{mmc}} $ are used to transform  the equation (\ref{constraint}) into:
\b \label{gaugecondition}  \partial^\top_\alpha\left[\partial^\top\cdot K-4(B\cdot x)^{-1} B\cdot K\right]+2 B\cdot x \partial^\top_\alpha \left[(B\cdot x)^{-2} B\cdot K\right]=0\, .\e
Let us express the Lagrangian in terms of the gauge field $K_\alpha=B^\top_\alpha \Phi_{\mathrm{mmc}}$ and the matter field $\psi$: 
$$ {\cal L}(\psi,K)= K\cdot Q_1K +K\cdot\partial^\top \partial^\top\cdot K+H\psi^\dag \gamma^0\left[ -\ii\barra{x}\gamma^\alpha \left(\partial_{\alpha}^\top+iK_\alpha\right)+2\ii\right] \psi\,.$$ 
Then the application of the constraint (\ref{gaugecondition}) results in the following gauge fixing Lagrangian:
$$ {\cal L}_{GF}(K)= c K\cdot\partial^\top\left[\partial^\top\cdot K-4(B\cdot x)^{-1} B\cdot K\right]+2c B\cdot x K\cdot \partial^\top \left[(B\cdot x)^{-2} B\cdot K\right],$$
where $c$ is the gauge fixing parameter. Hence, our model is equivalent to the Abelian gauge theory with a precise choice of the gauge $$\Lambda=-\frac{1}{2} (x\cdot B) ^{-2}= -\frac{H^2}{2} (B^\top\cdot B^\top) ^{-2}\, .$$ 

It is interesting to note that the vector field $K_\alpha\equiv B_{\alpha}^{\top}$ satisfies also the same field equations as the massless vector field (\ref{mlvfe}) without any constraint:
\b \label{simplf}
\left(Q_0-2\right)B_{\alpha}^{\top}+2x_\alpha
\partial^\top\cdot B^{\top}+H^{-2} \partial^\top_\alpha \partial\cdot B^{\top}=0\,.\e
Using the relation
$$\partial^\top_\beta\partial^{\top \beta}B_{\alpha}^{\top}=8x_\alpha B\cdot x+ 2B_{\alpha}^{\top},\;\; \partial^{\top }\cdot B^{\top}=4 x \cdot B,\;\; \partial^\top_\alpha \partial^{\top }\cdot B^{\top}= 4B_{\alpha}^{\top}\,.$$
one can show that the field equation (\ref{simplf}) is also invariant under the  gauge transformation:
$$B_{\alpha}^{\top} \rightarrow B_{\alpha}^{'\top}=B_{\alpha}^{\top}+\partial^\top_\alpha f(x),$$
where $f(x)$ is an arbitrary scalar field. In this case the Lagrangian ${\cal L}(\psi,\Phi_{\mathrm{mmc }},B^\top)$ becomes
\begin{equation*}
{\cal L} = \Phi_{\mathrm{mmc}} Q_0 \Phi_{\mathrm{mmc}}+H\psi^\dag \gamma^0\left[ -\ii \barra{x}\gamma^\alpha \left(\partial_{\alpha}^\top+\ii B^\top_\alpha \Phi_{\mathrm{mmc}}\right)+2\ii \right] \psi  + B^\top\cdot Q_1B^\top +B^\top\cdot\partial^\top \partial^\top\cdot B^\top\, . 
\end{equation*} 
The gauge fixing Lagrangian in this case becomes
$ {\cal L}_{GF}(B)= c_1 H^2 B \cdot B+c_2 \left(\partial^\top \cdot B\right)^2$ 
where $c_1$ and $c_2$ are the gauge fixing parameters. 

The meaning as a field on dS   of the projection $B^\top_\alpha$ of the constant  $B_\alpha$ is  encapsulated  in its behaviour under the 5-dimensional representation of the dS group \cite{gaha88}. Defining the projector $\Theta$ as $(\Theta B)_{\alpha}(x)= \theta_{\alpha}^{\beta}(x) B_{\beta}$, we have:
\begin{equation}
\label{BBT}
(\Theta R B)(x)= B^\top (R^{-1}x)\, . 
\end{equation}

  This field cannot be written in terms of the de Sitter plane waves and also $\partial^\top\cdot B^\top\neq 0$, therefore it cannot be associated with any elementary field in dS space. Also the norm and direction of $B^\top_\alpha$ everywhere in the de Sitter space-time are different for different coordinate system. This raises the question of its measurability. 
  
%  This constant vector field in five-dimensional space is very similar to the constant scalar field in four dimensions. But unlike the scalar field, the constant vector field has an arbitrary norm and arbitrary direction. This constant arbitrary direction in five-dimensional space, when projected on four-dimensional space-time, breaks the isotropic symmetry of the latter.

\section{One-loop corrections}
\label{onelc}
The advantage of our toy model is that it can be used to define the interaction between  scalar and spinor fields within the framework of the transformation (\ref{localtran}). Furthermore, it is free of any infrared divergence. It is important to note that at the null curvature limit the arbitrariness of $B_\alpha$ can be fixed completely by imposing the condition that the Yukawa interaction between scalar  and spinor fields be obtained. The arbitrariness of $A_\alpha$ in the two-point function could be removed from experimental results at the null curvature limit or from the interaction of the scalar field with the other fields in the tree level approximation \cite{ahjata}.

For our model the classical Lagrangian reads:
$$ {\cal L}(\psi,\Phi_{\mathrm{mmc }})= \Phi_{\mathrm{mmc}} Q_0 \Phi_{\mathrm{mmc}}+H\psi^\dag \gamma^0\left(-\ii  \barra{x}\barra{\partial}^\top+2\ii\right)\psi+\psi^\dag \gamma^0 \barra{x}\barra{B}^\top \Phi_{\mathrm{mmc}}\psi\, ,$$
which is invariant under the transformations (\ref{localtran}). The scalar two-point function is presented in (\ref{mth}) and the spinor two-point function is \cite{bagamota01}:
$$  S(x,x^{\prime})=\frac{\ii H^2}{8\pi^2}\frac{(\barra{x}-\barra{x}^{\prime})\gamma^4}{(1-{\cal Z}+\ii \tau\epsilon)^2}\,. $$
The scalar and spinor propagators $(G_{\phi}, G_{\psi})$ can be obtained easily from the above analytic two-point functions and the $\Phi-\psi-\psi$ vertex (${V}_{\Phi-\psi-\psi}=\psi^\dag \gamma^0 \barra{x} \barra{B}^\top \Phi_{\mathrm{mmc}}\psi$) comes from the interaction Lagrangian. It is important to note that the $B_\alpha$ is determined at the null curvature limit. Similarly to QED in the one-loop approximation there are three ultraviolet divergence diagrams, one-loop scalar propagator, one-loop spinor propagator, and one-loop vertex function. For the one-loop scalar propagator, we have two vertices in the points $x$ and $x^{\prime}$ and two spinor propagators $[G_\psi (x,x^{\prime})]^2$. The one-loop spinor propagator is constructed from two vertices,  one spinor propagator $(G_\psi (x,x^{\prime}))$,  and one scalar propagator ($G_{\phi}(x,x^{\prime})$). The one-loop vertex correction is constructed from three vertices at the points $x, x^{\prime}$ and $x^{\prime\prime}$ and three propagator $G_{\psi}(x,x^{\prime}), G_{\psi}(x^{\prime},x^{\prime\prime})$ and  $G_{\phi}(x^{\prime\prime},x)$. Since the quantum fields are constructed from the Bunch-Davies vacuum state, then the non-local terms are zero (there are not infrared divergent) and also the ultraviolet terms can be regularized at the null curvature limit exactly like it can be done  in Minkowski space-time.

\section{Conclusion and outlook}
\label{conclu}

The de Sitter ambient space formalism has permitted  to define a unique Bunch-Davies vacuum state for a quantum field theory that includes  \textit{mmc} scalar field and  linear quantum gravity. The infrared divergence is non-existent in either the quantization of the scalar field $\Phi_{\mathrm{mmc}}$, or the linear quantum gravity. In this construction the two-point functions are all analytic. Moreover, it is  clear that the \textit{mmc} scalar field may be viewed as similar to the Higgs field. By using this formalism one can also define the interaction between scalar and spinor fields  through  the transformation (\ref{localtran}) and the covariant derivative (\ref{gcd}). Now we have the all necessary building blocks for the construction of a unitary super-gravity in dS universe \cite{ta14} and unified theory of all interactions in de Sitter space. This will be considered in a forthcoming paper.

\vspace{0.5cm} 

{\bf{Acknowledgements}}:  The authors wish to express their particular thanks to Professor Jean Iliopoulos for pointing out the problem of gauge theory of our toy model. We are also grateful to Edouard Br\'ezin, Eric Huguet, Jean Iliopoulos, Richard Kerner, Salah Mehdi, Jacques Renaud, and Shahriar Rouhani, for helpful discussions.

\begin{appendix}

\section{Vacuum and ``one-particle'' states}
\label{vacone}

In the ambient space formalism, two solutions of the \textit{mcc} scalar field equation \eqref{mcceq} can be written in terms of the dS plane waves $(x\cdot \xi)^{-1}$ and $(x\cdot \xi)^{-2}$, where $\xi^\alpha$ lies in the positive cone $C^+=\left\{\xi^\alpha\in \R^5|\; \xi \cdot\xi=0, \; \xi^0>0 \right\}$ \cite{brgamo94,brmo96}. These solutions have simple or multiple singularities and must be regularised in a distributional sense.  Hence,  we consider their extensions to the complexified dS space-time \cite{brgamo94,brmo96}: 
$$ M_H^{(c)}=\left\{ z \in \C^5;\;\;\eta_{\alpha \beta}z^\alpha z^\beta=-H^{-2}\right\}$$
%\b =\left\{ (x,y)\in \R^5\times \R^5;\;\; x^2-y^2=-H^{-2},\; x\cdot y=0\right\}\,.\e
 Then one  uses the analytic complex dS plane waves to give a correct meaning to  the field operator   \cite{ta14} (for details see \cite{gagarota08} section VI):
\b \Phi_{\mathrm{mcc}}(z)=\sqrt{ c_0 }\int_{\mathbb{S}^3} \ud\mu({\bf \xi}) \left\lbrace\; a_{\mathrm{mcc}}({\bf\tilde{\xi}})(z\cdot\xi)^{-2}
+a_{\mathrm{mcc}}^\dag({\bf \xi})(z\cdot\xi)^{-1} \right\rbrace\, ,\e
where $\xi^\alpha=(1, \vec \xi, \xi^4)$, $\tilde \xi^\alpha=(1, -\vec \xi, \xi^4)$. Operators $a_{\mathrm{mcc}}$ and $a_{\mathrm{mcc}}^\dag$ obey the homogeneity conditions $a_{\mathrm{mcc}}(\lambda{\bf \xi})=\lambda^{-1} a_{\mathrm{mcc}}({\bf \xi})$ and  $a_{\mathrm{mcc}}^\dag(\lambda{\bf \xi})=\lambda^{-2} a_{\mathrm{mcc}}^\dag({\bf \xi})$ respectively.  The vacuum and the ``one-particle'' states for this field are defined as \cite{ta14}:
\b a_{\mathrm{mcc}}({\bf \xi})|\Omega\rangle=0\,,\;\; ;\;\;\;\;\; a_{\mathrm{mcc}}^\dag({\bf \xi})|\Omega\rangle=|1_\xi^c \rangle\, , \e 
with $$\langle 1_{\xi'}^c
 |1_\xi^c\rangle=\delta_{\mathbb{S}^3}(\xi-\xi'), \quad \int_{\mathbb{S}^3} \ud\mu({\bf \xi})\delta_{\mathbb{S}^3}(\xi-\xi') =1\,.$$ 
The vacuum state $|\Omega\rangle$ in this case is exactly equivalent to the Bunch-Davies vacuum state \cite{brgamo94,brmo96}.  

The analytic two-point function in terms of complex dS plane waves reads \cite{brgamo94,brmo96}:
\b \label{tpfscinint} W_{\mathrm{mcc}}(z,z')=\left<\Omega|\Phi_{\mathrm{mcc}}(z)\Phi_{\mathrm{mcc}}(z')|\Omega\right>=c_0\int_{\mathbb{S}^3}\ud\mu(\xi) (z\cdot\xi)^{-2}(z'\cdot\xi)^{-1}\,,\e 
where $c_0$ is obtained by using the local Hadamard condition. One can easily calculate (\ref{tpfscinint}) in terms of the generalized Legendre function \cite{brmo96}:
\b \label{atpfc} W_{\mathrm{mcc}}(z,z')=-\frac{\mathrm{i} H^2}{2^4\pi^2} P_{-1}^{(5)}(H^2 z\cdot z')=-
\frac{H^2}{8\pi^2}\frac{1}{1-{\cal Z}(z,z')}=\frac{H^2}{4\pi^2}(z-z')^{-2}\,. \e
 The Wightman two-point function ${\cal W}_{\mathrm{mcc}}(x,x')$ is the boundary value (in the distribution sense, according to Theorem A.2 in \cite{brmo96}) of the function $W_{\mathrm{mcc}}(z, z')$  \cite{brmo96}. 
 
Apart from the polarization constant five-vector $A_\alpha$, the \textit{mmc} field operator in the complex dS space-time can be defined properly from the quantum field operator of the \textit{mcc}  field and the identity (\ref{mmcmcc}):
\begin{equation}
\label{phimmczz}
\begin{split}
\Phi_{\mathrm{mmc}}(z;A)&=\sqrt{ c_0 }  \int_{\mathbb{S}^3} d\mu({\xi}) \left\lbrace\; a_{\mathrm{mmc}}(
{\bf \tilde{\xi}};A)\left[-2(A\cdot\xi)(z\cdot\xi)^{-3} \right] +\right.\\
& \left. a_{\mathrm{mmc}}^{\dag}(
{\xi};A)\left[-(A\cdot\xi)(z\cdot\xi)^{-2} + (A\cdot z)(z\cdot\xi)^{-1}\right]
\; \right\rbrace\, .
\end{split}
\end{equation}
 The vacuum and the ``one-particle'' states in this case are defined as:
\b  a_{\mathrm{mmc}}({\bf \xi};A)|\Omega\rangle=0,\;\; a_{\mathrm{mmc}}(\lambda{\bf \xi};A)=\lambda^{-1} a_{\mathrm{mmc}}({\bf \xi};A), $$ $$ a_{\mathrm{mmc}}^\dag({\bf \xi};A)|\Omega\rangle=|1_\xi^{\mathrm{mmc}};A \rangle,\;\; a_{\mathrm{mmc}}^\dag(\lambda{\bf \xi};A)=\lambda^{-2} a_{\mathrm{mmc}}^\dag({\bf \xi};A),$$
$$ \langle 1_{\xi'}^{\mathrm{mmc}};A
 |1_\xi^{\mathrm{mmc}};A\rangle=\delta_{\mathbb{S}^3}(\xi-\xi')\, .\e
The vacuum state is unique. With these definitions of the field operator, the vacuum, the ``one-particle'' states, and the two-point function (\ref{mth}) can be reconstructed. The ``one-particle'' state $|1_\xi^{\mathrm{mmc}},A \rangle$ does not transform under a unitary irreducible representation of the dS group.

 The field operator for the choice (\ref{zpolar}) is: 
$$ \Phi_{\mathrm{mmc}}(z) =\sqrt{ c_0 } \sum_{l=0}^4 \int_{\mathbb{S}^3} d\mu({\xi}) \left\lbrace\; a_{\mathrm{mmc}}(
{\bf \tilde{\xi}};A^{(l)})\left[-2(A^{(l)}\cdot\xi)(z\cdot\xi)^{-3} \right] \right.$$ $$\left.+a_{\mathrm{mmc}}^{\dag}(
{\xi};A^{(l)})\left[-(A^{(l)}\cdot\xi)(z\cdot\xi)^{-2} + (A^{(l)}\cdot z)(z\cdot\xi)^{-1}\right]
\; \right\rbrace\, .$$
In this case, we have:
\b \langle 1_{\xi'}^{\mathrm{mmc}};A^{(l)}
 |1_\xi^{\mathrm{mmc}};A^{(l')}\rangle=\eta^{ll'}\delta_{\mathbb{S}^3}(\xi-\xi')\, .\e

The field operator for the choice (\ref{so4inv})  can be  simply obtained with the choice $A_\alpha\equiv (1,0,0,0,0)$ in the equations (\ref{phimmczz}) and we have:
\b \langle 1_{\xi'}^{\mathrm{mmc}}
 |1_\xi^{\mathrm{mmc}}\rangle=\delta_{\mathbb{S}^3}(\xi-\xi')\, .\e

It is important to note that the massless minimally coupled scalar field on Bunch-Davies vacuum state may be a composite field \cite{ellis17}, {\it i.e.} it is not an elementary scalar field. For an arbitrary $A$, the massless minimally coupled scalar field does not transform under a UIR of the dS group \cite{gasiyou10}. On the quantum level, the quantum states, which depend on the constant vector field $A$ may be associated to the ``soft particle'' in a gauge theory \cite{strom17} and we have:
$$ A\neq A' \Longrightarrow \langle 1_{\xi'}^{\mathrm{mmc}};A
 |1_\xi^{\mathrm{mmc}};A'\rangle=0\, .$$
 
In the above discussion the arbitrariness of $A$ correspond to the one-particle state, which is called the first perspective. There is another perspective for the construction the quantum state in which the arbitrariness of $A$ corresponds to the different vacuum state:
 \b |\Omega,A\rangle\, .\e
We recall that with this $A$-dependence,  the vacuum state is not fully de Sitter invariant. It is just O$(4)$ invariant. 

%In this paper, our choice of  the first perspective allows us to  obtain a unique vacuum state $|\Omega\rangle$ for QFT in dS space, for which the theory is free of any infrared divergence and is analytic. Analyticity is crucial for the loop calculation in the interaction case. Moreover we think that our approach might help in  the construction of a quantum gravity without infrared divergence and a unitary super-gravity in dS space, an essential step in any attempt to establish a unified field theory.

\end{appendix}

\end{document}